# The Design of Tangible Digital Musical Instruments

**Gareth W. Young**

Dept. of Computer Science / Music

University College Cork

`G.Young@cs.ucc.ie`

**Katie Crowley**

Trinity College Institute of Neuroscience

Trinity College Dublin

`K.Crowley@tcd.ie`

**Abstract**   Here we present guidelines that highlight the impact of haptic feedback upon the experiences of computer musicians using Digital Musical Instruments (DMIs). In this context, haptic feedback offers a tangible, bi-directional exchange between a musician and a DMI. We propose that by adhering to and exploring these guidelines the application of haptic feedback can enhance and augment the physical and affective experiences of a musician in interactions with these devices. It has been previously indicated that in the design of haptic DMIs, the experiences and expectations of a musician must be considered for the creation of tangible DMIs and that haptic feedback can be used to address the physical-digital divide that currently exists between users of such instruments.

## 1.  Introduction

General advances in technology have always influenced the field of music technology, facilitating the modern musician's requirement for new devices and encouraging creative expression. Music technology has a deep-rooted history of performance and a close-knit relationship with human interactions with musical devices. Through the use of natural sound resonating objects, such as reeds, bells, pipes, and others, humans have made possible the creation of musical instruments. Traditionally, it was the limitations of the sound-generating device that determined the design of an instrument. However, this fact has never deterred the making of music from most any man-made or naturally resonant object.

With the discovery and widespread application of electricity in the early part of the $20^{th}$ century, the number of new mediums for sound generation increased. The majority of early electronic musical instruments were keyboard based, drawing upon the universal success of acoustic instruments such as the piano and harpsichord. Notable exceptions that deviated from this design principle are instruments such as the Theremin [1] and other instruments that made use of gesture sensitive inputs to control timbre, pitch, and/or volume. Another example would be the Trautonium [2], which operated via a touch sensitive strip across its length. Of the keyboard-based instruments, advancements in functionality were achieved via increasing the devices sensitivity or manipulation through the development of additional knobs and buttons, for example, the Ondes Martenot [3] and the Electronic Sackbutt [4].

Digital Musical Instruments (DMIs) have very few limits and the potential design possibilities are vast. Beyond musical performance, in devices that operate on a one-to-one interaction, new DMIs are also encompassing other jurisdictions of musical composition. Artists may become proficient in the use of a singular instrument or they may choose to become the master of a multi-instrumental controller. Musicians may concentrate all of their efforts into increasing their skill in playing a stand-alone instrument or they may choose to master the control of multiple sound sources through digital manipulation. A performer may also have an indirect influence over an installation or live recital, becoming a unique and often difficult to control aspect of a performance. Beyond the musician, performance itself has also changed. The musical medium is also no longer a static performance, as a single musician or ensemble on stage, it moves beyond this. It can be inclusive of the movements of a dancer, a dance troupe, and even the audience itself. The inclusion of multiple free movements into music production paves the foundations for a more expansive interaction.

## 2. Background

In computer music, virtually any gesture can be captured and translated into a control signal. In the application of DMIs, these gestures are often used as a control source for complex sound synthesis modules. With the separation of interface from sound source, new musical devices are afforded near endless freedom of form. However, they are becoming unrecognisable, as the gestures captured by a device do not require resemblance of anything ever applied before. The multiple combinations of different styles of interface design have protracted the performance techniques that musicians are afforded in performance. This is indicative in the increased popularity of DMIs in contemporary music, as they have been embraced and accepted as a new means for artistic expression.

Musical interface models based upon the playing principles of musical instruments have previously been seen [5]. In Figure 1, we present our own concept of a closed-loop model of a tangible musical instrument. It is proposed here that if a DMI wishes to be considered as tangible, a number of steps must be followed to ensure this. In this context, both artist and instrument can be observed as two separate entities that are independent of each other. The link between user and instrument mediates between the minor components contained within. The relationship between these two modules is realised through gestures made and gestures captured. The musician or artist is independently providing the intention (often attained through training and previous experience) and the necessary gestures specific to the interface. The instrument captures these physical interactions and processes them into a form of control data. The sound generator makes use of the data collected from gestures captured by applying control parameters to a physical sound generating design. The physical separation of these modules is impossible to achieve in acoustic instruments (represented in Figure 1, as the gesture interface is rarely removed from the sound source. DMIs allow us to separate the user from the instrument, permitting us to rethink the relationships formed between the two. For example, a gesture can be made and the sound generated varies in some way; however, the gesture does not necessarily have to relate to a control change in the sound generator, as it may also convey performance information that is not audible. What has become apparent from observing current DMI trends is that whilst performers have been given absolute freedom of gesture capture, they have at the same time eliminated haptic feedback, a key feedback channel of information through which they can measure the response of the instrument and the accuracy of their movements. In the realm of gesture capture, synthesis algorithms and control rate data have been separated from the sound producing mechanisms along with the performer. The capture of human performance with such devices forces the user to rely heavily on the proprioceptive, visual, and aural data cues, or more simply put: "...the computer music performer has lost touch" [6].

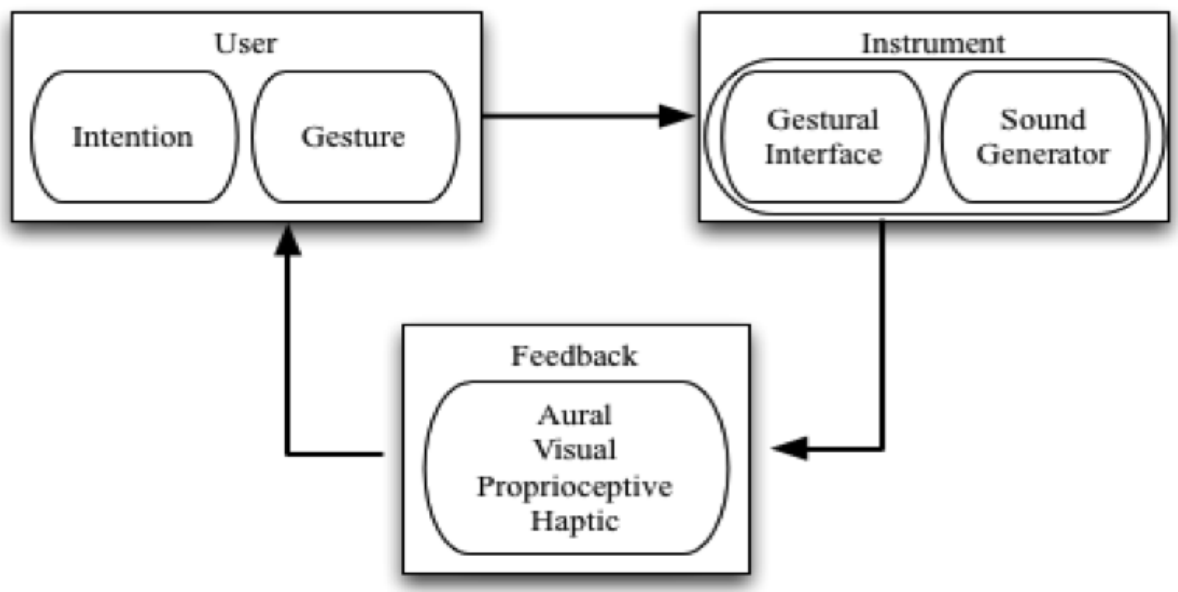

**Figure 1:** Closed Feedback Loop of a Tangible Musical Instrument.

## 3. Design Guidelines

Through the amalgamation of digital music technology and electronic musical instruments, DMIs have emerged. A DMI is a musical instrument that is capable of producing sound via digital means. They are specifically constructed with a separable control interface and sound generator; however, these are not always separate. The mapping of a gestural interface to a sound generator translates the input gestures into sound control signals that are applied to a sound generator. The separation of these two elements enables musicians to approach the creation of music differently from how they would with an acoustic instrument, as the physical constraints of sound generation and input gesture are no longer inseparable. This approach allows for the sonification of gestures or the creation of a sound-generating algorithm that is controlled via an unknown or undefined input gesture.

DMI designs, such as the Rhythm'n'Shoes [7], The Sound Flinger [8], the Haptic Carillon [9], The Vibrobyte [10], StickMusic [11], and The Plank [12] have demonstrated the successful application of haptic feedback in musical devices. However, the majority of commercial interfaces in the field of digital synthesis have focussed on simulating the effects of acoustic keyboard instruments (such as the piano, harpsichord, or organ) and bowed instruments (such as the violin, viola or cello) [13]. Furthermore, previous research has highlighted that many of these DMIs fail to balance complexity with usability and that they lose transparency due to the separation of sound generator and gestural interface [14]. In the guidelines outlined herein, haptic information that can be used to address these issues will be focused on and will attempt to resolve problematic issues of interaction. We therefore propose consideration of the following guidelines when creating a tangible DMI:

- Transparent in use.
- Reactive and communicative to as many of the user's senses as possible.
- A tangible interpretation of the device's reaction must be possible.
- Clarity of affordances delivered via the sensor technologies applied.
- Consistency in the information displayed to the user.
- The application of clear and stable mapping methodologies.
- Clear and consistent constraints for the interpretation of gestures made.

### 3.1.  Transparent

It must be possible to clearly determine the function of the instrument, for both the musician and the observing audience, as it is easier to recognise an action than to recall one. The musician should be able to quickly and easily identify the options offered by the instrument, and how to access them. In contrast, options that are "out of sight" can be difficult to find and their functionality unintuitive.

### 3.2.  Reactive

In relation to a devices transparency, information must be presented to as many of the user's senses as is possible in a timely and logical manner to emphasise the effect of the input interaction upon the system in use (see Figure 1). The provision of feedback gives the user confirmation that an action has been performed successfully, or otherwise. Physical or auditory activational feedback can be used to indicate to the musician whether or not their intended action has successfully been performed.

### 3.3.  Tangible

All information related to the system's reaction should be presented to the musician clearly and they should also be able to interpret meaning easily, this will serve to enhance discoverability and improve the musician's overall understanding of the device.

### 3.4.  Clarity of Affordance

A DMI that is designed with familiar features should be done so with clarity in how these features react and should therefore respond in a recognisable and familiar way. For DMI designers, affordance means that as soon as the musician sees the instrument, they should have some knowledge as to how to use it. Users need to be able to tell how to access the full spectrum of options available to them. When the affordances of a physical object are perceptually obvious it is easy to know how to interact with it. The standard example used for explaining affordance is the door handle. A round doorknob invites you to turn the knob. A flat plate invites you to push on the plate to push the door outwards. A graspable handle invites you to pull on the handle to pull the door open towards you.

### 3.5. Consistent

The location, appearance, significance, and behaviour of an interface must be consistent for it to be effectively learned. In achieving this, when errors are made the interface will allow musicians to recover and continue without any additional mental or physical strain. Good design principles suggest that the same action has to cause the same reaction, every time. This is particularly important in the context of a DMI where the musician can experience flow [15] during the performance and requires the certainty and assurance that their intended action will be achieved, without having to focus on interface-related concerns. Consistency is also important in terms of learning how to play and master the DMI. People learn by discovering patterns and this is particularly relevant in a musical context. Consistency is key to helping users recognise and apply patterns. Conversely, inconsistency can causes confusion and uncertainty because something does not work in the way the user expects. Forcing users to memorise exceptions to patterns and rules increases the cognitive burden and leads to distraction, particularly during a performance. Consistency is important for instilling confidence in both the system and instrument.

### 3.6. Clear and Stable Mapping

The mapping of gestures in a spatial context and the systems temporal responses should be clear and stable. Controls should be positioned in logical and intuitive ways, for example, it is obvious that a slider control designed to manipulate volume maps the direction of "up" to increase volume and "down" to decrease. Nonconventional mappings need to be learned and can conflict with consistencies guidelines, however, they are permissible when an appropriate or valid reason exists.

### 3.7. Constraints

The introduction of physically identifiable, logical, and clear limitations upon an interaction will prevent errors and assist in interaction interpretation by both the musician and the system in use. Interfaces must be designed with restrictions so that the system can never enter into an invalid state - the same principle applies to DMI design. Constraints can be physical, such as a size limitation on an object or restriction threshold on the angle of movement of a lever, for example.

## 4. Discussion

It is hoped that through the application of these design guidelines that advances in the field of Computer Music will be made. Specifically, it is foreseen that the study of interactions between performers and digital instruments in a variety of contexts will continue to be of interest in this field far beyond the scope presented here. Further research on digital musical instruments and interfaces for musical expression should continue to explore the role of haptics, previous user experience, and the frameworks that are constructed to quantify the relationship between musical performers and new musical instruments. The complexities of these relationships are further compounded by the skills of musicians and are far more meaningful than a physically stimulating interaction and should therefore be explored further.

The designer of a DMI is often the performer and a DMI may take many forms; from concept to performance tool. In a similar vein, in the design processes of computer interfaces, evaluation tools are applied iteratively, in cycles that address the design issues raised within the previous sequence. An example of this can be seen in Norman's Seven Stages of Action as a design aid in interaction design [16]. Whilst appraising a DMI, the musician must constantly questions certain aspects of usability when applied to specific tasks. For example:

- Can I achieve my goals accurately and effectively?
- Am I working productively and efficiently?
- Is the device functioning as I expect it to?
- At what rate am I acquiring new skills?

Emergent DMI systems require further measures for an accurate appraisal of the user's experience when applying the device in a musical context. In a traditional HCI analysis, a device is evaluated in a specific context and the evaluation methods are expert-based heuristic evaluations or user-based experimental evaluations. Only by determining context is it possible to interpret correctly the data gathered. Therefore, it is suggested that to fully understand the effects of tangibility upon DMI performance, specific evaluation techniques must be formulated, such as the application of functionality, usability, and user experience evaluation methods [17].

The ideas presented in this paper have only begun to explore the possibilities of tangibility in future DMI designs. The ideas presented endeavoured to present thoughts towards the influence of feedback on a user's perception DMI tangibility. Beyond this, future research goals will include the development of laboratory tools that will assist in the creation of a DMI design environments that will allow designers to experiment with different communication paradigms and gestural interface models. Within this space, composers, performers, and DMI designers will be able to explore the affordances of new sensor technologies in the creation of new instruments for musical expression.

## 5. Conclusion

By following the guidelines presented here, haptically enabled DMI designs will be fully communicative to all senses and present computer musicians with an array of carefully designed tools for their own artistic endeavours. Furthermore, we believe that the experiences of the audience will also be improved upon as clarity between the musician's actions and the system's response will be achieved.

# References

[1] Theremin World, "What's a Theremin?" (December, 2005). [Online]. Available: http://www.thereminworld.com/Article/14232/whats-a-theremin

[2] New York Times, "ELECTRICITY, ETHER AND INSTRUMENTS; Some Considerations, Reflections and Inferences Regarding the Modern Cult of Vitamineless Art and the Synthetic Esthetic" (6 September 1931). [Online]. Available: http://timesmachine.nytimes.com/timesmachine/1931/09/06/96381853.html

[3] T. Bloch, "Ondes Martenot" (12 March 2015). [Online]. Available: http://www.thomasbloch.net/en_ondes-martenot.html

[4] Canada Science and Technology Museums Corporation, "HUGH LE CAINE *ELECTRONIC SACKBUT* SYNTHESIZER," (27 February 2015). [Online]. Available: http://cstmuseum.techno-science.ca/en/collection-research/artifact-hugh-le-caine-electronic-\sackbut-synthesizer.php

[5] J. Rovan & V. Hayward, "Typology of Tactile Sounds and their Synthesis in Gesture-Driven Computer Music Performance," in *Trends in Gestural Control of Music*, M. Wanderly & M. Battier, Eds., Editions IRCAM, 2000.

[6] N. Castagné, C. Cadoz, J. L. Florens & A. Luciani, "Haptics in computer music: a paradigm shift" EuroHaptics, 5-7 June 2004.

[7] S. Papetti, M. Civolani & F. Fontana, "Rhythm'n'shoes: a wearable foot tapping interface with audio-tactile feedback" in *New Interfaces for Musical Expression*, Oslo, 2011.

[8] C. Carlson, E. Marschner & H. McCurry, "The Sound Flinger: a Haptic Spatializer" in *New Interfaces for Musical Expression*, Oslo, 2011.

[9] M. Havryliv, F. Naghdy, G. Schiemer & H. T., "Haptic Carillon – Analysis & Design of the Carillon Mechanism" in *New Interfaces for Musical Expression*, Pittsburgh, 2009.

[10] K. Mcdonald, D. Kouttron, C. Bahn, J. Braasch & P. Oliveros, "The Vibrobyte: a Haptic Interface for Co-Located Performance," in *New Interfaces for Musical Expression*, Pittsburgh, 2009.

[11] H. C. Steiner, "StickMusic: Using haptic feedback with a phase vocoder" in *New Interfaces for Musical Expression*, Hamamatsu, 2004.

[12] B. Verplank, M. Gurevich & M. Mathews, "The Plank: Designing a simple haptic controller," in *New Interfaces for Musical Expression*, Singapore, 2002.

[13] E. R. Miranda & W. M. Wanderley., "Haptic Music Controllers: Tactile and Force Feedback," in *New Digital Musical Instruments: Control and Interaction Beyond the Keyboard*, vol. 21, Midleton, WI: A-R Editions, 2006, pp. 71-83.

[14] T. Murray-Browne, D. Mainstone, N. Bryan-Kinns & M. D. Plumbley, "The medium is the message: Composing instruments and performance mappings," in *New Interfaces for Musical Expression*, Oslo, 2011.

[15] M. Csikszentmihalyi, & I. S. Csikszentmihalyi, (Eds.), "Optimal experience: Psychological studies of flow in consciousness," Cambridge University Press, 1992.

[16] D. A. Norman, "The design of everyday things: Revised and expanded edition," New York: Basic books, 2013.

[17] G. W. Young & D. Murphy, "HCI Models for Digital Musical Instruments: Methodologies for Rigorous Testing of Digital Musical Instruments," in the *Int. Symp. on Computer Music Multidisciplinary Research*, Plymouth, UK, 2015.